\pdfoutput=1
\documentclass[journal=apchd5,manuscript=letter]{achemso}
\setkeys{acs}{doi = true}
\usepackage[version=3]{mhchem} 
\usepackage{xcolor}

\author{Andrew Lininger}
\email{arl92@case.edu}

\author{Michael Hinczewski}
\email{mxh605@case.edu}

\author{Giuseppe Strangi}
\affiliation{Case Western Reserve University, Department of Physics 2076 Adelbert Road, Cleveland, OH, USA 44106}
\alsoaffiliation[CNR-NANOTEC]
{CNR NANOTEC, Istituto di Nanotecnologia and Department of Physics, University of Calabria, Rende, Calabria, IT 87036}
\email{gxs284@case.edu}

\phone{+12165076952}

\title[]
  {General Inverse Design of Thin-Film Metamaterials With Convolutional Neural Networks}

\abbreviations{CNN,ID}
\keywords{Machine Learning, Deep Learning, Photonic Design, CNN, Optical Coatings, Thin Film Photonics}

\begin{document}

\begin{abstract}
The design of metamaterials which support unique optical responses is the basis for most thin-film nanophotonics applications. In practice this inverse design problem can be difficult to solve systematically due to the large design parameter space associated with general multi-layered systems. We apply convolutional neural networks, a subset of deep machine learning, as a tool to solve this inverse design problem for metamaterials composed of stacks of thin films. We demonstrate the remarkable ability of neural networks to probe the large global design space (up to $10^{12}$ possible parameter combinations) and resolve all relationships between metamaterial structure and corresponding ellipsometric and reflectance / transmittance spectra. The applicability of the approach is further expanded to include the inverse design of synthetic engineered spectra in general design scenarios. Furthermore, this approach is compared with traditional optimization methods. We find an increase in the relative optimization efficiency of the networks with increasing total layer number, revealing the advantage of the machine learning approach in many-layered systems where traditional methods become impractical.
\end{abstract}


\vspace{1em}
Nanostructured metamaterials have become ubiquitous in modern photonics, providing a platform for engineered control over light-matter interaction. Metamaterials operate by modifying the sub-wavelength spatial distribution of materials in a deliberate pattern, creating optical responses which go beyond those found in natural materials.\cite{maccaferri2019,Lui2011,capasso2014} In particular thin-film metamaterials, metamaterials with varied material composition restricted to a single axis, have garnered widespread attention in the photonics community and are commonly utilized in applications such as optical coatings and sensors.\cite{elkabbash2021,Sreekanth2019}

The engineered design of metamaterials sustaining a desired optical response in a given spectral range, termed inverse design (ID), is the basis for most optical applications. Successful metamaterial design requires the exploration of the global materials parameter space to engineer an optimized response. \cite{Yao2019,Molesky2018,Peurifoy2018,pilozzi2018,Liu2018} This approach is counter to traditional electromagnetic simulation methods, where electromagnetic theory is applied to calculate the spectral response of a known material structure.\cite{gallinet2015} This presents a major problem for nanophotonic ID, since in general analytic methods to calculate material structures from a given spectral response are not known. Furthermore, the underlying correlations between spectra and structure are often complex and are difficult to generalize for systems with many independent layers of different materials.

\begin{figure}[t]
    \centering
    \includegraphics[width=1.0\linewidth]{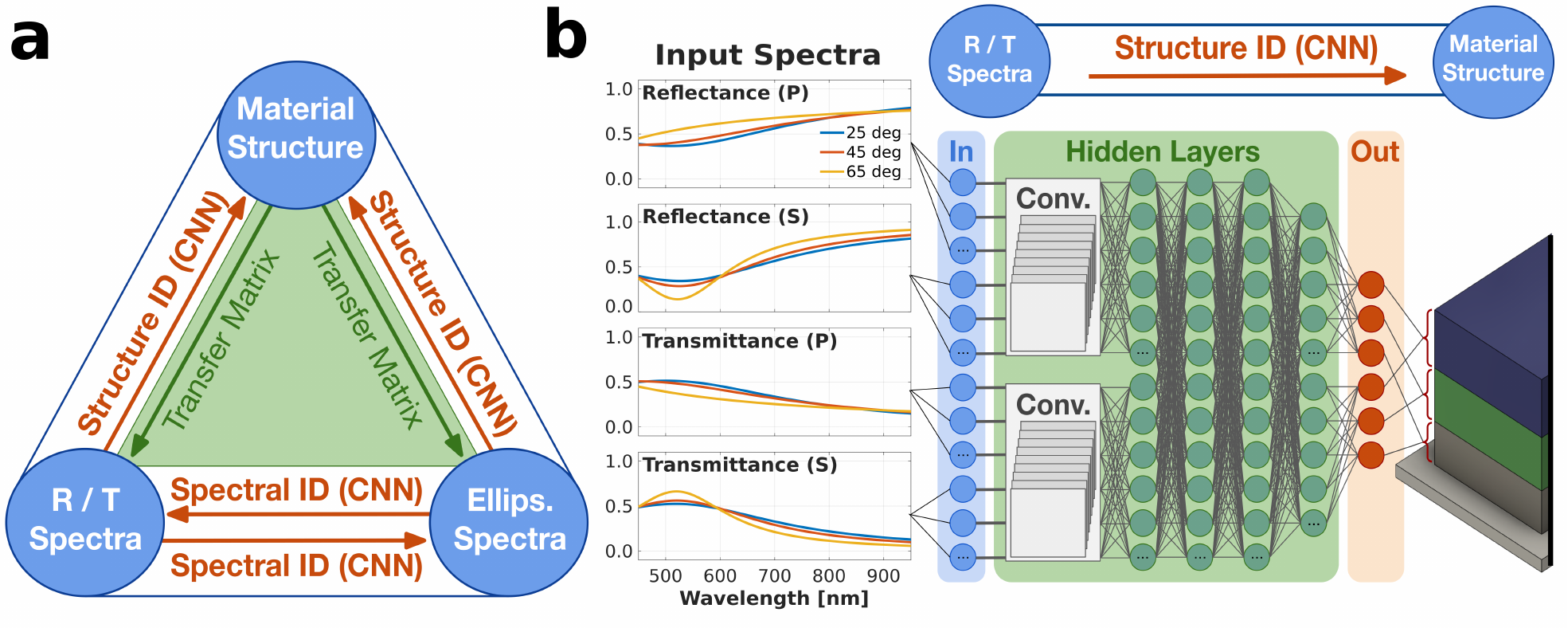}
    \caption{Illustration of the full inverse design problem and convolutional neural network (CNN) solution. \textbf{a} The inverse design triangle for optical metamaterials consists of reflectance / transmittance spectra, ellipsometry spectra, and metamaterial structure. CNN based inverse design legs for metamaterial structure (Structure ID) and spectral response (Spectral ID) are shown in orange, while the completely determined transfer matrix legs are shown in green. \textbf{b} Pictorial representation of the CNN network structure in the reflectance / transmittance structure ID problem.}
    \label{fig1}
\end{figure}

Deep neural networks have been previously implemented as a tool for photonic ID.\cite{Zhou2019,Ma2020,So2020} Machine learning based approaches to photonic ID are typically proposed to replace traditional gradient based methods, which can often be computationally expensive to implement. The capabilities of deep neural networks result from their ability to learn complex patterns by generalizing large quantities of data through the training of a large number of internal weight parameters. \cite{HORNIK1989} These networks have shown the remarkable ability to solve the ID problem accurately and efficiently in specific nanophotonic systems such as thin-films,\cite{Liu2018,Kim2020,Jiang2020,Kaya2019,Qu2019,liu2021} 2D metasurfaces,\cite{Malkiel2018,Jiang2019,Christensen2020,Ma2018,Nadell2019,Balin2019,Kojima2021,yeung2021} and core-shell nanoparticles,\cite{Peurifoy2018,Akashil2020,kim2021} within a limited parameter space. Furthermore, neural network models have been implemented to replace the typical forward electromagnetic methods used to simulate optical systems.\cite{Peurifoy2018,Li2019} In these studies significant advances in solution time have been demonstrated with respect to traditional simulation methods.

Machine learning approaches to solving the photonic ID problem can come in a variety of types: deep neural networks\cite{Liu2018,Peurifoy2018,Li2019,Akashil2020,Kojima2021}, convolutional neural networks\cite{Kim2020,Lin2020,Sajedian2019,Asano2018,so2019-1}, and generative models\cite{yeung2021,Christensen2020}, among others. Furthermore, machine learning methods have been applied for inverse design to predict both continuous and discrete photonic design parameters.\cite{QIU2021}

Convolutional neural networks (CNNs) are a subset of deep neural networks which utilize convolutional layers to assess aspects of an input signal.\cite{Lawrence1997,Krizhevsky2017,LIU2017} This approach uses local connectivity in the convolutional layers, whereby the input characteristics are evaluated in an internal representation space allowing interpretation of spectral features by their relative proximity. The stacking of subsequent layers forms filters which allows for evaluation of an input signal on both local and global scales. The convolutional approach allows for greater generalization ability and predictive accuracy in a wide range of computational tasks when compared to traditional deep neural network methods. \cite{LeCun2015,Moncau2018}

In this work, we apply CNNs to the inverse design of an important class of metamaterials---planar multilayer stacks where each individual film is much thinner than the wavelength of incident light. CNNs have been successfully applied to other specific photonic ID problems~\cite{Lin2020,Sajedian2019,Asano2018,so2019-1}, however these problems  are typically within a relatively constrained set of possible structures. Here we explore the utility of CNNs from a more general perspective, expanding upon these previous works by applying CNN techniques to a larger and more general library of available materials and ranges of layer thicknesses, effectively searching a combinatorially large global design space (up to $\sim$ 10$^{12}$ possible parameter combinations). This library does not make use of any particular design schema (i.e. periodic photonic crystal structures) in predicting photonic response, including all combinations of materials and thicknesses for each independent layer in the stack. This makes the implementation more similar to the true inverse design of a general forward simulation method.

The overarching goal of the ID problem is to predict sets of structural parameters that achieve certain design targets. In this regard, we consider the structure -- spectral correspondence in terms of both the reflectance / transmittance and ellipsometric spectral responses.  Depending on the target and the desired output, the general ID problem can be broken down into several individual sub-problems, illustrating the interrelationships between a typical structure and multiple spectral representations. We show that each of these problems, including the inverse design of photonic structures from both reflectance / transmittance and ellipsometric structures, respectively, and the conversion between independent spectral types, can be solved by implementing an independently trained CNN model.  We describe these in detail in the next section.

\section{Results and Discussion}

\subsection{Structure of the Inverse Design Problem}

We conceive of the ID problem as threefold, shown in Fig.~\ref{fig1}a. All relationships between the three representations of thin-film metamaterials---material structure, ellipsometric spectra and reflectance / transmittance spectra---consist of interconnected design problems. We seek to fully explore the ID problem for a given metamaterial structure by elucidating all relationships between the representations. 

Most thin-film optical engineering is represented by the structure ID problem, determining a specific stack of material layers which produces a particular spectral response.  This involves determining the composition and thickness of each layer, and the ordering of the layers in the stack.  In particular, most practical applications of metamaterials involve the structure ID of reflectance ($R$) / transmittance ($T$) spectra (left upward arrow in the Fig.~\ref{fig1}a triangle). We design CNNs to take all $R$ and $T$ spectra for both polarizations of incident light ($R_p$, $R_s$, $T_p$, $T_s$) at [25, 45, 65] degree incident angles and output individual layer parameters (material and thickness).  The range of incident angles used in the training can be easily adapted for specific applications.

Ellipsometry, another method of spectral analysis, provides optical information about metamaterial structure and composite materials by taking into account the phase relations in polarized reflected light.\cite{aspnes1981} Ellipsometry is a standard experimental method in nanophotonics, yielding two spectral variables ($\Psi$ and $\Delta$). $\Psi$ relates to the ratio in magnitude of the p- and s- polarized reflectance Fresnel coefficients while $\Delta$ relates to the phase shift between the same coefficients, where:
\begin{equation}
    \frac{r_p}{r_s} = \tan(\Psi)e^{i \Delta}
\end{equation}

The analysis of metamaterials based on structure ID of $\Psi$ and $\Delta$ spectra is the primary purpose of most commercial ellipsometry software. However, the traditional methods of model fitting employed can be difficult due to generally requiring detailed prerequisite knowledge about the target structure including layer thicknesses and composite materials.  In contrast, our CNN-based ellipsometric structure ID (right upward arrow in the Fig.~\ref{fig1}a triangle) attempts to solve this problem without such constraints, with a search occurring over the entire global design space.  The designed CNN takes as input $\Psi$ and $\Delta$ spectra at [25, 45, 65] degree incident angles and outputs individual layer parameters.  Note that the reverse of the structure ID problem left/right downward arrows in the Fig.~\ref{fig1}a triangle) is generally straightforward:  if the system structure is completely known, both ellipsometry and reflectance / transmittance spectra are completely determined and can be easily calculated with the transfer matrix method.\cite{Chilwell1984} 

Finally, the third leg of the ID problem is the ability to reconstruct all ellipsometry spectra ($\Psi$, $\Delta$) from the complete reflectance / transmittance spectra ($R_p$, $R_s$, $T_p$, $T_s$), and vice versa, for nanophotonic structures residing in the design space (bottom arrows in the Fig.~\ref{fig1}a triangle). This spectral ID problem is non-trivial since it requires the reconstruction of phase or transmittance data, respectively, and can in principle be degenerate (multiple possible solutions in our global design space) without detailed knowledge of the underlying system structure.  Despite these complexities, we show that these CNNs we can tackle this aspect of the problem as well.

Throughout the text we will be using the terminology introduced here---structure ID for the upward arrows in the triangle, whose output is material structure, and spectral ID for the bottom arrows, where the output is a certain optical spectrum---to describe the different CNNs we developed.  To define a concrete design space, we trained our CNNs on sample data from thin-film metamaterials of 1-5 layers, with layer thicknesses from 1 to 60 nm, and with a set of possible materials:  Ag, Al$_2$O$_3$, ITO, Au and TiO$_2$.  (Full details of the training, as well as the network structure of the CNNs, can be found in the Methods.)  We treat the training/testing of each total layer number (1-5) subspace as a separate problem, so there is a different network trained for each layer number.  Excluding degenerate cases where consecutive layers are the same material, the design space amounts to $\sim 10^{12}$ possible parameter combinations for the most complex case (for 5 layers, with thickness at 1 nm resolution, see Supporting Information for more discussion on the design parameter space).  However the approach we present is not limited to this particular range of structural parameters and choice of material library. The training can readily be adapted to a different design space depending on the specific photonic problems of interest by altering the training dataset.

\begin{figure}[t]
    \centering
    \includegraphics[width=1.0\linewidth]{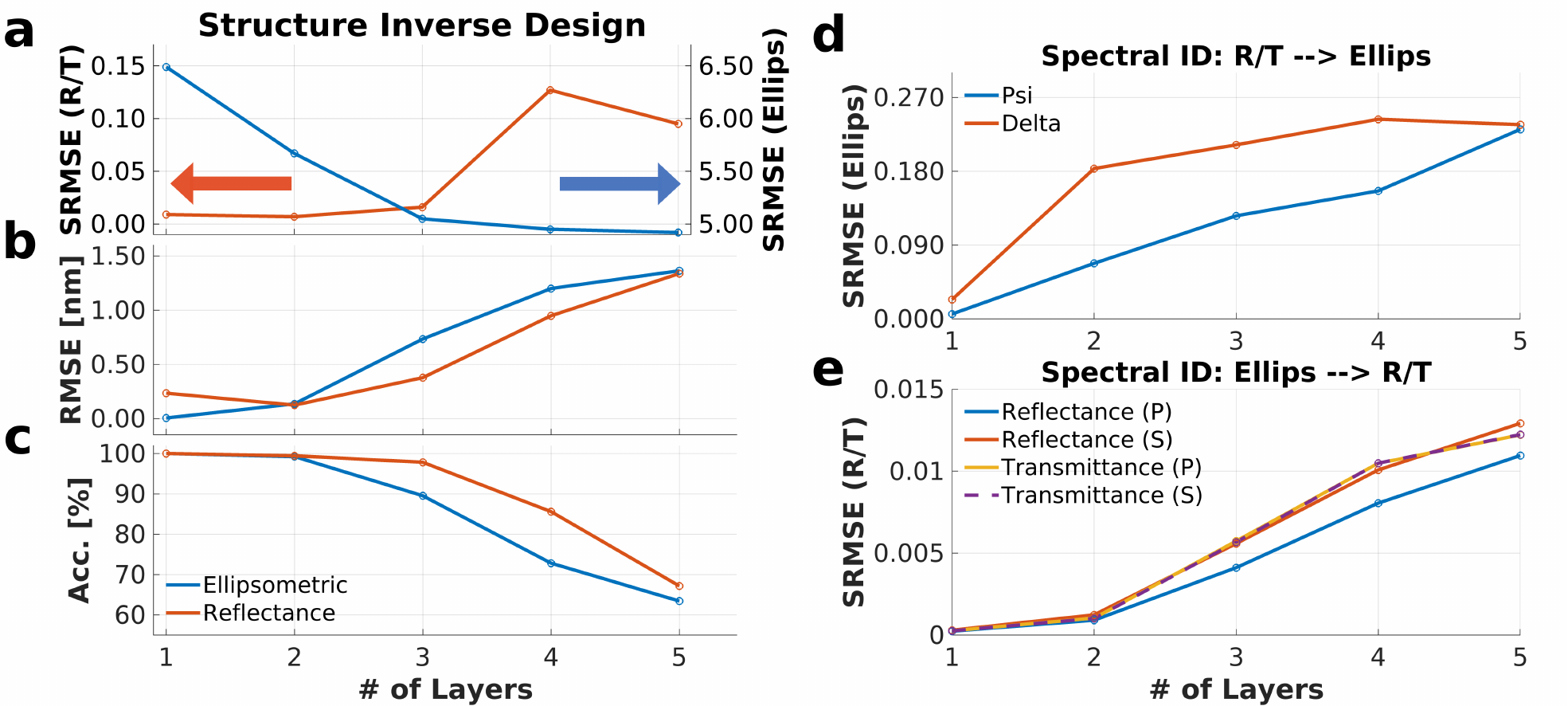}
    \caption{Performance metrics for the inverse design convolutional neural networks (CNNs). \textbf{a-c} Performance of the structure ID CNNs for both ellipsometry spectra (blue) and reflectance / transmittance spectra (red) as a function of total layer number. Networks are evaluated from an independent test dataset. Specific metrics are shown: \textbf{a} spectral RMSE between the input spectra and spectral response of the output structure averaged between all distinct spectral sub-types (ellipsometric structure ID ($\Psi$, $\Delta$)[deg] and reflectance / transmittance structure ID ($R_p$, $R_s$, $T_p$, $T_s$) [unit-less]), \textbf{b} average layer thickness RMSE [nm], and \textbf{c} average layer material accuracy [\%]. \textbf{d-e} Network performance for spectral ID in the design of both \textbf{d} ellipsometric spectra from reflectance / transmittance spectra ($\Psi$ and $\Delta$ output, [deg]) and \textbf{e} reflectance / transmittance spectra from ellipsometric spectra ($R_p$, $R_s$, $T_p$ and $T_s$ output, [unit-less]) spectra as a function of total layer number.}
    \label{fig2}
\end{figure}

\subsection{Prediction of Material Spectral Response}

There are multiple ways of evaluating the effectiveness of structure and spectral ID CNNs.  As a first step, we focus on the calculated spectral response from the network output structural predictions.  In the case of the structure ID CNNs, whose output is a set of structural parameters (layer thicknesses, compositions), the calculated spectral response consists of the reflectance / transmission or ellipsometric ($\Psi$ and $\Delta$) spectra of the output structure, calculated using the transfer matrix method.  These spectra can be compared to the input spectra that were design targets, and a root mean squared error (RMSE) calculated over the spectral range of interest (450 -- 950 nm) and for all angles([25, 45, 65] degrees).  We refer to this metric as the spectral RMSE, coming in two types (reflectance / transmittance [unit-less] or ellipsometric [deg]) depending on the spectral type used. For the case of the spectral ID problem, we can also formulate a spectral RMSE metric, by comparing the output of the network (which in this case is directly a spectrum) to the ground truth from the system producing the input spectrum.  All the evaluation results in this section and the subsequent ones are based on a testing set of systems consisting of previously unseen examples, drawn from the same design space as the training set (see Methods for details).

Spectral RMSE is in many cases the most practical measure of ID performance, particularly in situations with degenerate solutions---different structural parameters that produce similar spectra.  For example the target spectrum in the testing set may have been produced by a certain structure, but the network can find an alternative structure (different materials, different thicknesses) that yields a closely matching spectrum (and hence small spectral RMSE).  This would still be a valid solution of the ID problem, fulfilling the intended spectral design task, however the predicted structure does not match the target design structure. Ideally we would like to minimize the spectral RMSE, while matching the target design structure. Balancing these two imperatives is a goal of our approach.

The average spectral RMSEs in the structure ID problem for both ellipsometric and reflectance / transmittance spectra are shown in Fig.~\ref{fig2}a, evaluated over an independent test dataset.  To understand the scale of the spectral RMSEs, note that the total output range for reflectance ($R$) and transmittance ($T$) is from 0 to 1, and for the ellipsometric variables ($\Psi$ and $\Delta$) this range is from 0 to 90 degrees and 0 to 360 degrees, respectively.  For reflectance / transmittance spectra, observed spectral RMSE is low (less than $1\%$ of the total spectral output range) and increases with increasing total layer number to a maximum of $13\%$ of the output range for 4 layered systems. Ellipsometric spectral RMSE is observed to decrease with increasing total layer number, from $3\%$ of the spectral output range to $2\%$ for 1 and 5 layered systems, respectively. Tolerances for acceptable spectral RMSE are necessarily application dependent, however we consider these results good, especially for the systems with fewer layers. Notably, the effectiveness of the CNNs extends even to systems with many layers, where the design space is far larger. This is all the more remarkable because spectral RMSE is not used directly as part of the loss function for CNN training (see Methods for details of the loss function), and hence the performance in this regard is a byproduct rather than an explicit goal of the training process. This is an important feature of our implementation of the CNN methods, since we consider the training of networks to predict the underlying structure a major goal. This overcomes a deficiency in some traditional methods which solely minimize the spectral RMSE in the optimization loss, since these methods can be especially prone to encountering degeneracy within the design space. This can lead to solutions which locally minimize the spectral RMSE, but do not typically select the correct structure.


Spectral RMSE results for the spectral ID CNNs are shown in Fig.~\ref{fig2}d-e as a function of total layer number. For both spectral ID problems, the spectral RMSE is observed to increase with increasing total layer number. This increase is expected with the exponentially increasing size of the parameter space, however the CNN responses maintain a low RMSE even for relatively high total layer number systems. The maximum average spectral RMSE occurs for 5 layer systems with $0.1\%$ and $1\%$ of the total output range for ellipsometric [deg] and reflectance / transmittance [unit-less] spectral types, respectively). This result demonstrates the ability of CNN models to accurately correlate spectral types for general classes of real systems. This result is impressive, since general relationships between reflectance / transmittance and ellipsometric spectra are not well defined analytically. It is notable that the CNN models perform comparably for both p and s polarization in transmittance. This is explained by relatively low transmittance for most systems above three layers, given that an average layer thickness of $30$ nm and three layers significantly increases the probability of optically opaque structures.

We note that training a network to solve the spectral ID problem directly (correlating between the two distinct spectral types, following the bottom right arrow of the triangle) performs better than the indirect approach of first solving the structure ID problem from $R$, $T$ to material parameters, and then using the transfer matrix to get the $\Psi$, $\Delta$ spectra (structure ID spectral RMSE, the alternative path of following the left upward and right downward arrows).  Comparing Figs.~\ref{fig2}a and e, the spectral RMSEs are typically an order of magnitude better for the direct path. This underlines the importance of having separate trained networks for all three legs of the ID triangle. Even though spectral ID yields smaller spectral RMSE values, in many design applications structure ID is essential, because we desire to explicitly predict the material parameters.  We explore this point further in the next section.

\subsection{Prediction of Material Structure Parameters} 

For the case of structure ID, since the network output consists of structural / material parameters, we can use these to formulate an alternative measure of CNN performance:  how well does the network predict the correct material in each layer, and how well does it predict each layer thickness?  These metrics are imperfect in the sense that by construction the predictive accuracy should tend to decrease as the layer number increases.  This is because for larger layer numbers there may be many degenerate structures with similar optical response.  But it is still interesting to measure the impact of this degeneracy issue on predictive performance.  The results are shown in Fig.~\ref{fig2}b-c, as a function of total layer number. In both reflectance / transmittance and ellipsometric structure ID, trained CNN models are able to reproduce the correct materials and thickness for the corresponding input spectra with high precision ($>90\%$ materials accuracy, $<1$ nm thickness RMSE) in up to three total layers, averaged over the entire test dataset (see Supporting Information for more details on the network response to test data). For systems with higher total layer number the CNN models are still able predict structures with significantly correct materials and thicknesses. However, a decrease is observed in the material accuracy and consequently and increase in thickness RMSE for large total layer number systems. This decrease is attributed both to construction: the expected increase in degeneracy for much larger parameter spaces; and to network performance: a decreased ability of the CNN to generalize the greater complexity of higher order systems. In general, this vast increase in parameter space greatly increases the demand upon the CNN for generalization of the spectral response.

The comparison with spectral RMSE is instructive: as seen in Fig.~\ref{fig2}a, reflectance / transmittance spectral RMSE increases by $0.11$ and ellipsometric spectral RMSE decreases by $1.57$ deg, with decreasing average layer material accuracy and increasing average layer thickness RMSE. The comparison of these two metrics is an indication of degeneracy within the included parameter space, since the CNN can more often come to predictions with a similar spectral response while utilizing physical structures with less correspondence to the target structure.

\begin{figure}[t]
    \centering
    \includegraphics[width=1.0\linewidth]{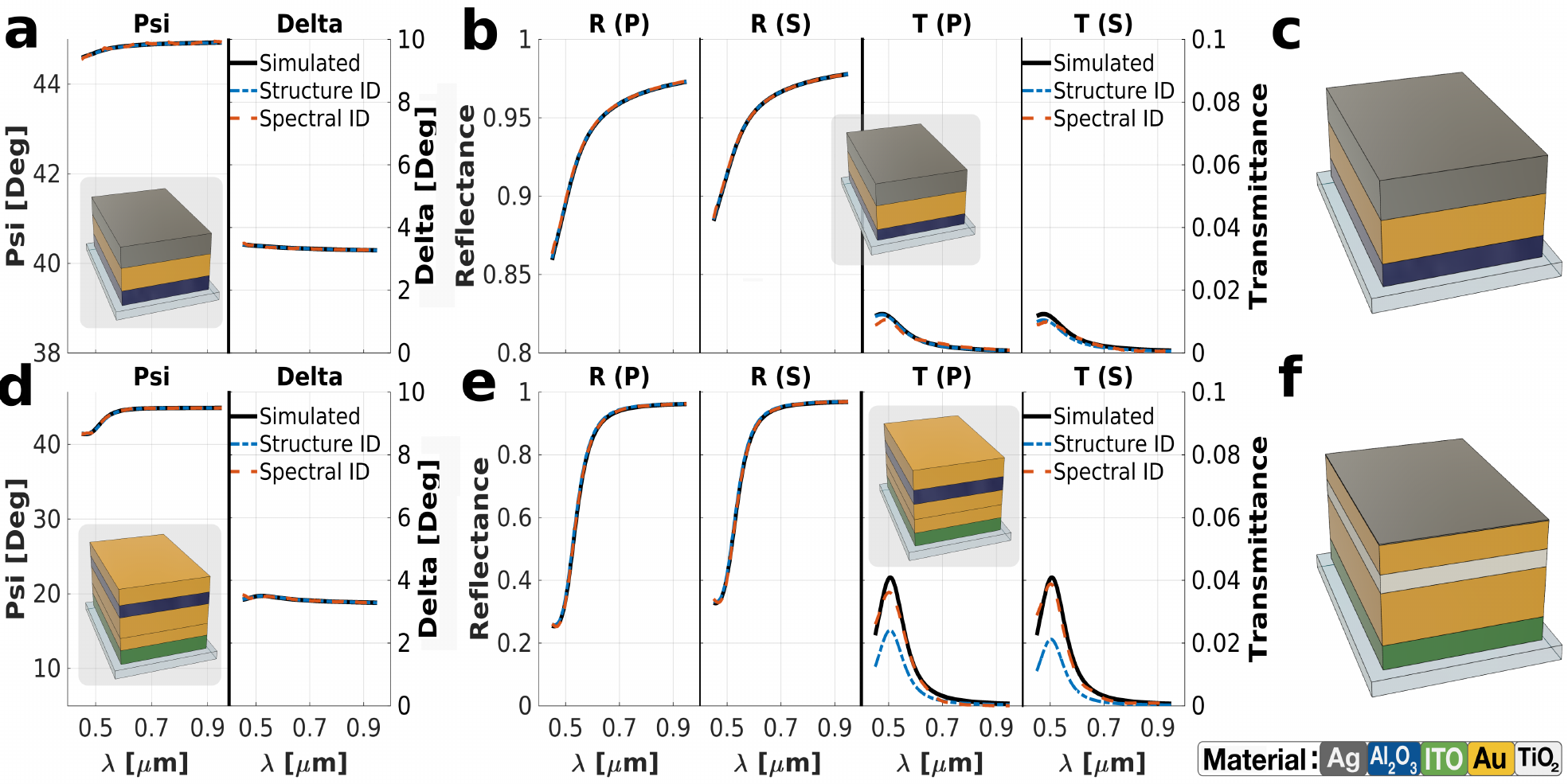}
    \caption{Examples of convolutional neural network (CNN) inverse design for structures within the training parameter space. Inverse design of \textbf{a-c} ground truth structure 1 shown in \textbf{c}, \textbf{d-f} ground truth structure 2 shown in \textbf{f}. Structures are drawn to scale with material colors in the figure legend. Simulated spectra (black) are plotted against structure inverse design (blue) and spectral inverse design (red) CNN prediction results. \textbf{a} Ellipsometry inverse design of structure 1, predicted structure is shown in the inset. \textbf{b} Reflectance / Transmittance inverse design of structure 1, predicted structure is shown in the inset. \textbf{d} Ellipsometry inverse design of structure 2, predicted structure is shown in the inset. \textbf{e} Reflectance / Transmittance inverse design of structure 2, predicted structure is shown in the inset.}
    \label{fig3}
\end{figure}

\subsection{Inverse Design of In-Library Structures}

Two examples of structure ID and spectral ID performed with CNNs are shown in Fig.~\ref{fig3}, plotted against the corresponding input spectra. Both the target design structure and inverse designed structures are shown. Spectral comparisons are computed by operating directly with the transfer matrix code on the CNN structure output. In each case both the ellipsometric and reflectance/transmittance CNNs predict a structure which reproduces the spectral response to a high degree of accuracy. Remarkably, the CNN model is able to predict spectral response irrespective of sharp local features in the input spectra, as can be seen for both the relatively sharp and broad features (see Supporting Information for more discussion on this point). This feature of the CNN prediction is not equally represented in the spectral RMSE metric, and is affected by both the CNN feature extraction approach and an operation on a limited library of materials.

An example 3-layer system is shown in Fig.~\ref{fig3}a-c, and the CNN model is seen to predict the correct materials subspace corresponding to the input spectra. Because of this, any error in the spectral response is then a result of fine-tuning individual layer thicknesses to optimize the spectra. For the 5-layer system shown in Fig.~\ref{fig3}d-f, the CNN does not predict the correct materials subspace for the ground truth structure. However, for both ellipsometric and reflectance/transmittance spectra, spectral response of the chosen structure closely matches that of the target spectra to within $0.1$ deg and $2\%$ R/T maximum error, respectively. This is an interesting property of the CNN based inverse design, that incorrect structures are often predicted which still produce a spectral response that closely mimics the target spectra. The existence of such structures is a known result of the spectral degeneracy for many layered systems, compounded by the finite CNN accuracy as discussed above.\cite{Liu2018} However, the solution types shown in Fig.~\ref{fig3}d-f reside in material subspaces which have been specifically excluded from the training data set because of the repeating Au layers. This means the CNN was not shown any examples of this type previous to the prediction of this structure. That the CNN can come to such solutions is a positive feature of the ID process, indicative of generalization to a set of physically realistic principles encoded within the CNN. 

Spectral ID results for the presented systems are shown on the same axes as the corresponding input spectra in Fig.~\ref{fig3}. In each case the respective spectral ID CNNs accurately predict the corresponding spectra given the physical input. This is clear evidence of generalization in the CNN, since the underlying relationship corresponds only to a correlation based on physical systems and is, in principle, highly degenerate with no clear analytic mapping between spectral types.

\subsection{Creative Inverse Design of Synthetic Spectra}

\begin{figure}[t]
    \centering
    \includegraphics[width=1.0\linewidth]{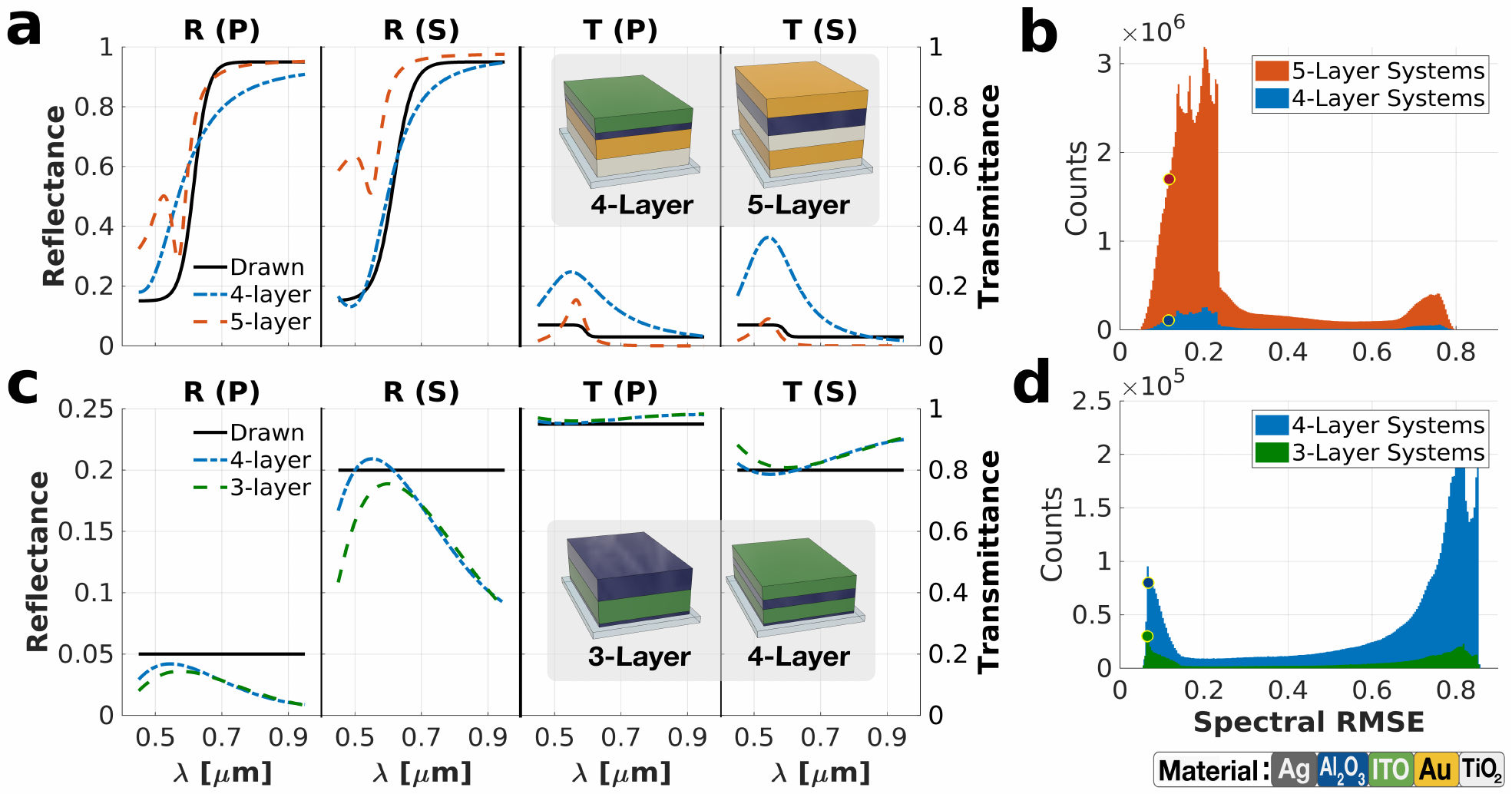}
    \caption{Examples of convolutional neural network inverse design for structures outside the training parameter space with a-b for the first spectra and \textbf{c-d} for the second spectra. \textbf{a} Drawn spectra (black) are plotted against the 4-layer (blue) and 5-layer (red) CNN prediction results, with predicted structures inset. Structures are drawn to scale with material colors in the figure legend. \textbf{b} Histogram of spectral RMSE for all possible solutions to the inverse design problem within the design parameter space. The CNN predicted results are shown as plotted points on the histogram for comparison. \textbf{c} Drawn spectra (black) are plotted against the 4-layer (blue) and 3-layer (green) CNN prediction results, with predicted structures inset. \textbf{d} Histogram of spectral RMSE for all possible solutions to the inverse design problem within the design parameter space.}
    \label{fig4}
\end{figure}

In practical applications of structure ID, the desired spectral response for a multi-layer thin-film metamaterial is generally not exactly represented by a system contained in the design parameter space probed by the training library, since it is possible to conceive of arbitrary spectral features in the target design. In these cases, CNN structure predictions are purely a generalization of the input spectra to similar features from the in-library parameter space, as the CNN typically predicts outputs in this space. Two examples are provided in Fig.~\ref{fig4}, designed to illustrate the usefulness of this approach in practical design scenarios. The purpose of the CNN models implemented in this study is the exploration of a large, general parameter space. In this space, the availability of certain spectral responses is generally limited by the finite size of the parameter space. Thus the correspondence of the calculated spectral response from the network prediction with the design target will necessarily be limited. The true measure of network performance is not the absolute spectral RMSE in this case, but a comparison of this value with spectral responses from structures within the allowed parameter space, which is discussed below.

In Fig.~\ref{fig4}a-b the input spectra mimic a reflectance filter with sharp cut-off at an arbitrary cut-off wavelength. This input spectra is not generated from any known metamaterial structure and is simply drawn from an appropriately scaled hyperbolic tangent function. Spectra corresponding to the predicted 4-layer and 5-layer structure ID CNN predictions are shown plotted against the drawn input spectra. In each case spectra from the predicted structures closely recreate the sharp edge in the reflectance and suppressed transmittance with non-negligible differences in the remainder of the spectra. The inability of the CNN to completely capture the input spectra is due to a limitation of possible structures in the design parameter space probed by the CNN, and not a deficiency in the predictive ability of the network. This is represented in Fig.~\ref{fig4}b, showing for each total layer number a histogram of spectral RMSE between the input structure and all possible structures within the design parameter space. Comparison spectra were obtained from structures with design parameters spanning the entire design parameter space, for a total of more than $6 \times 10^{10}$ and $3 \times 10^{13}$ structures (sampled at 0.5 nm thickness intervals) for 4 and 5 total layer number systems, respectively. The CNN predicted structures show a spectral RMSE in the top $95\%$ and $90\%$ of all possible structures for the 4-layer and 5-layer systems, respectively. These results are well above median, indicating the ability of the CNN to predict globally optimized structures. The prediction of very low spectral RMSE solutions is constrained by the non-existence of exact solutions in the finite parameter space. This is shown by the non-zero minimum in the spectral RMSE histogram. In general, completely matching solutions to drawn (not physically generated) spectra do not exist within this space. Also note that the spectral RMSE metric does not equally represent sharp features in the target spectra to which the CNN may have a modified response, as discussed above. Optimization of these features is varied and can be better evaluated on a case-to-case basis in line with the specific design objectives.

The input spectra in Fig.~\ref{fig4}c-d are also drawn and not generated from a physical system. These are meant to represent a broadband anti-reflective coating for the glass substrate. Spectral responses corresponding to structures predicted by the 3-layer and 4-layer CNNs are shown. In each case the CNN predicts a structure with spectral characteristics similar to the input, although the CNN predicted structure is not able to reproduce the completely flat spectral response. As discussed above, a histogram of spectral RMSE with possible spectral responses from all structures within the parameter space is shown in Fig.~\ref{fig4}d. In this example the CNN produces highly optimized structures, with spectral RMSE in the top $98\%$ and $99\%$ of all possible structures for the 3-layer and 4-layer systems, respectively. The predicted structures show spectral behavior similar in magnitude to the input spectra, and evaluation of the full space histogram shows that spectral errors are a limitation of the design space and not of the CNN optimization.

Notably, both of these examples feature arbitrarily engineered spectra, in the sense that they are drawn from appropriately scaled mathematical functions and not generated from any physical structure. This demonstrates the usefulness of CNN based structure ID in general design scenarios, and the ability of CNNs to produce globally optimized structures in a given parameter space.

\subsection{Comparison with Other Methods}

In response to the complexity of photonic structure ID, the optimized computational design of nanophotonic structures traditionally relies on optimization techniques employing forward electromagnetic solvers.\cite{Molesky2018} Generally the performance of these traditional methods in producing globally optimal solutions are limited by both the volume of parameter space being probed and the efficiency of generating forward solutions. Since optical spectra for thin-film nanophotonic systems can be generated at relatively low computational cost via the transfer matrix method, the comparison with traditional methods of optimization is an important insight into the practicality of CNNs as a design tool for general nanophotonic systems.\cite{Chilwell1984}

Comparisons were made with least squares (Levenberg-Marquardt)\cite{Anzengruber2012} and genetic algorithms,\cite{Storn1997,Froemming2009} two common methods utilized in structure ID of nanophotonic systems. Since the general ID problem consists of a continuous basis of individual layer thickness within discrete material subspaces, the least squares algorithm was globalized by performing a brute force random global search of all material subspaces. The genetic algorithm is able to natively accommodate the discrete basis. All comparison algorithms were written in Python with common modules, optimized for evaluation time and evaluated on the same high performance computational resources as the CNN (see the Methods section for full details).

\begin{figure}[t]
    \centering
    \includegraphics[width=1.0\linewidth]{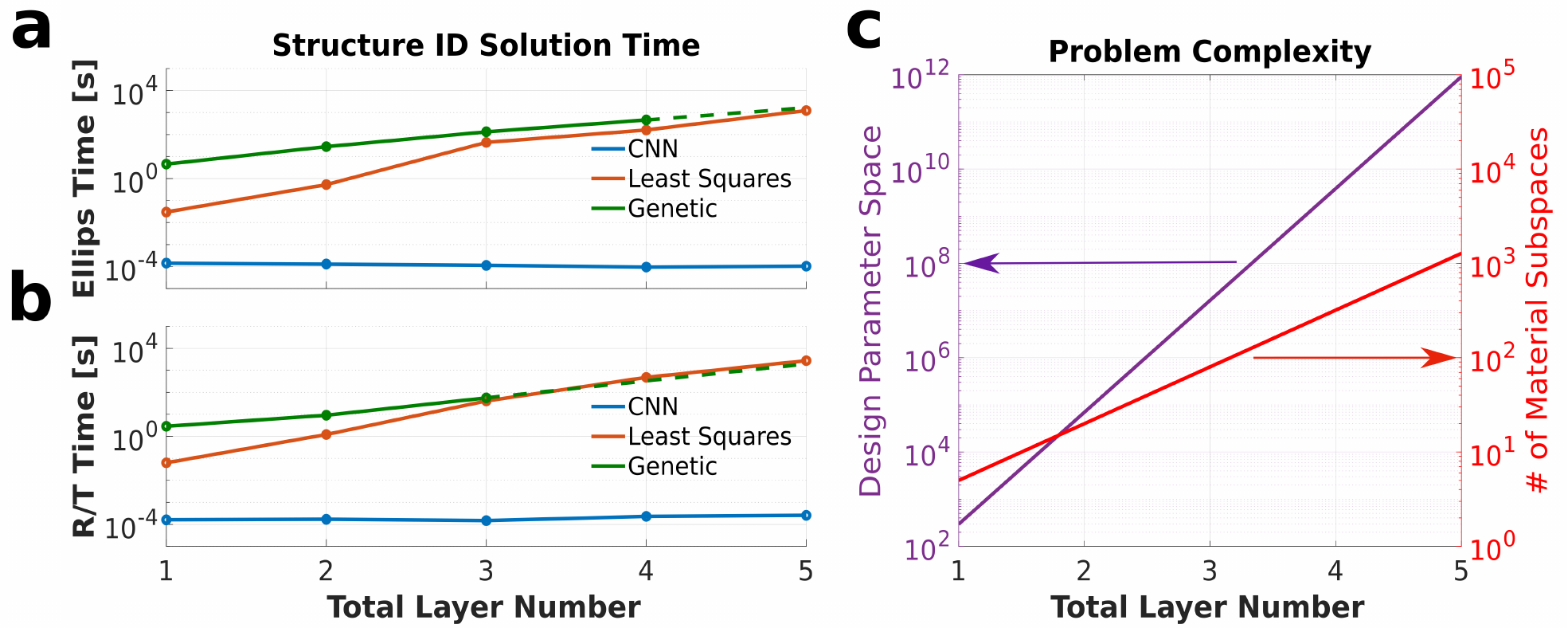}
    \caption{Timing comparison results for structure ID convolutional neural networks and comparable optimization techniques.  \textbf{a} Ellipsometric inverse design solution time (s) for the CNN (blue), least squares optimization (red) and genetic algorithms (green) as a function of total layer number. Dashed lines indicate projected results based on an exponential regression of observed results. \textbf{b} Similar results for reflectance / transmittance spectra inverse design as a function of total layer number. \textbf{c} Complexity (purple) of the general inverse design problem (size of the probed parameter space, discritized by 1 nm) as a function of total layer number. Total number of allowed materials choices are also shown, as number of distinct material subspaces (red). These values grow exponentially with layer number, complicating the inverse design problem.
}
    \label{fig5}
\end{figure}

The observed solution times for both comparison methods and each system total layer number are shown in Fig.~\ref{fig5}a-b, along with the corresponding structure ID CNN solution time.  For systems with many layers, some results have been predicted from an exponential regression of the observed solution times. These are indicated by dashed lines in Fig.~\ref{fig5}a-b.  CNN training time has not been included in this analysis, since use of the CNN is in practice separated from training. Furthermore, the CNN models can be evaluated for many systems once the initial training has been completed, while the comparison methods require new initialization for each system. For every total layer number, in both structure ID of reflectance / transmittance and ellipsometric spectra, the CNN solution time is faster than the comparable optimization techniques by several orders of magnitude. 

Furthermore, the trend in system solution time is toward exponential growth with increasing total layer number for both of the comparison optimization techniques. The CNN solution time remains constant within an order of magnitude regardless of total layer number. This trend highlights the growing impracticality of conventional blind optimization solutions for general nanophotonic systems with high total layer number or high level of complexity. For five total layers least squares solution time is already 6 orders of magnitude greater than the CNN based optimization for both spectral types. This analysis accounts for inverse design accuracy by modifying the hyperparameters associated with each comparison method to minimize solution time while maintaining a layer materials accuracy and layer thickness RMSE similar to the CNN model prediction capabilities.

The exponential increase in solution time for the traditional optimization is a result of the increase in total parameter space size as shown in Fig.~\ref{fig5}c. This is a product of the number of material subspaces (all permutations of available material combinations) with the allowed range of thicknesses for each layer (counted by discretizing the space in 1 nm thickness intervals). Although further optimizations of both methods are potentially possible, the increasing exponential trend clearly favors CNN efficiency when designing systems with higher total layer number. Globally optimal solutions require traditional optimization techniques to repeatedly probe the entire design space, which becomes increasingly costly as the size increases. However, CNNs provide a fundamentally different approach by modeling the space in a fixed number of pre-trained weight parameters, so the solution time at evaluation is dependent only upon the number of nodes contained in the model irrespective of the volume of underlying design space. This difference in method is reflected in the observed improvements in solution time for CNNs as compared to these traditional methods.  

The field of design in thin-film photonics has seen the introduction of many methods beyond those mentioned above. In particular traditional techniques such as the needle method have been shown to produce thin-film photonic structures with extremely high fidelity to defined design targets.\cite{Tikhonravov1996,Tikhonravov2007,Sullivan1996} To illustrate the abilities of our method with respect to this class of design algorithms, a comparison was made between a numerical implementation of the needle method and our CNN models, over the test dataset described above. The results of this comparison can be found in the Supporting Information. From this analysis, it can be concluded that while the numerical needle method does work to minimize the spectral RMSE in predicted structures, the resulting predictions are not typically correct with respect to the known structural parameters. In fact for 4 and 5 total layers, the needle method predicts the correct layer material less accurately than random chance, while producing a spectral RMSE much lower than random chance for the parameter space. The CNN methods, however, produce structures with both optimized spectral RMSE and structural parameters. This is a positive feature of our method, that designing the CNN models to maximize the accuracy of predicted structural parameters results in a corresponding decrease in the spectral RMSE. This is counter to traditional methods which just attempt to minimize the spectral RMSE, and are much more prone to degenerate solutions.

In addition, many machine learning methods could in principle be substituted for the CNN methods implemented in this study since neural networks are essentially 'black-box' methods with different internal representations. The comparison between different network methods is relevant since the performance metrics shown above are directly influenced by the network representation efficiency.\cite{Kahn2020,He2016} To illustrate the ability of our CNN methods compared to other machine learning methods, two key comparisons have been considered. The results of these comparisons are shown in the Supporting Information. The two methods considered are more traditional fully-connected deep neural networks (FC-DNNs) and more recent ResNets. From this analysis, it is shown that our CNN method outperforms the FC-DNN method and performs similarly to the ResNet method (within $6\%$ spectral RMSE) for the reflectance / transmittance structure inverse design problem. This indicates that our CNN implementation performs well when compared to the breadth of possible single network machine learning methods for this specific problem.

\section{Conclusions}

We demonstrate the potential of convolutional neutral networks to solve the ID problem in thin film metamaterials for a completely general library, with multiple choices for both individual layer materials and thicknesses in systems with various total layer number. This problem is difficult due to the large input parameter space for systems with many layers. Furthermore, this convolutional machine learning approach is shown to solve the ID problem for all legs of the inverse design triangle. This includes the inverse design of physical structures based on both ellipsometric and reflectance / transmittance spectra individually, as well as the ability to translate between both spectral types for systems in the design parameter space. Using this method, inverse design is systematically applied to both real and synthetic spectra, allowing for the creative design of real physical systems based on arbitrarily drawn spectra. To illustrate the benefits of the machine learning approach to inverse design, these methods are then compared directly with common traditional methods of inverse design. The convolutional machine learning approach allows us to globally probe the design parameter space for a given library of materials and thicknesses, which can be difficult and costly with traditional methods. This illustrates the full generalizing ability of neural networks to produce systems with a desired spectral response for a wide range of input design parameters. 

The neural network based inverse design methods applied here are general in terms of the utilized design parameter library, in the sense that the choice of a specific design parameter space is not essential to the observed success of the method. The methods presented can be easily translated into specific inverse design problems based on the desired ranges of materials and thicknesses in the output structures. This is a possible extension of these results into practical implementations for real-life design scenarios. Another possible extension of this work includes increasing the possible range of spectral response by the addition of 2-D metamaterial structures into the existing framework. To overcome the major increase in parameter space associated with general 2-D structures as well as the burden of generating large numbers of structures and associated spectra, effective medium approximations of constrained structures could potentially be employed. Furthermore the role of machine learning in nanophotonics continues to be driven by the discovery and application of new techniques. The application of developing machine learning techniques always has the potential to increase the inverse design efficiency in future models.

\section{Methods}
\subsection{Generation of the Training Dataset}

Training data for the CNNs was generated using the transfer matrix method for layered thin-film metamaterial structures.\cite{Chilwell1984} Samples were generated with total layer number in the range of 1 – 5 discrete material layers. Layer thicknesses were chosen uniformly in the continuous range of [1, 60] nm, with materials chosen from a library of [Ag, Al$_2$O$_3$, ITO, Au and TIO$_2$]. Degeneracy is common problem in photonic inverse design, and an inevitable feature of a large design library, due to the fact that multiple structures can have very similar spectral response. To limit degeneracy in the structural parameters, possible structures with consecutive layers of the same material were removed in the training dataset. Our method still achieves high material accuracy and low layer thickness RMSE despite any remaining degeneracy, in comparison to the needle method.

All structures are simulated on an infinite glass substrate. Spectral response was calculated for 200 equally spaced points in the range of [450, 950] nm at the incident angles [25,45,65] deg. A training dataset of 200,000 sample structures was generated for each ID problem type, to be used in training the neural network models (see Supporting Information for more information on material optical properties and the generated dataset).  This corresponds to a sampling rate of only about 600 examples per material subspace for four-layer systems and about 150 examples per material subspace for five layer systems.

\subsection{Design and Tuning of the Convolutional Structure}
The neural network models discussed in this work generally employ a convolutional architecture. This consists of a series of 1-D convolutional layers followed by downsampling with a max pooling layer. The convolutional layers operating on the input spectral types independently. This is followed by a series of several fully connected deep layers which are fully connected to the output nodes. All deep layers are Relu activation. Mixing of information from different spectral types is accomplished in deep fully connected layers following the convolutional layers by adding together parallel layers. Dropout regularization is included following each hidden layer in the model. Raw spectra are passed directly to the network input without further manipulation, except for dividing the ellipsometric spectra by 45 deg to better range the input values.

A sample network structure can be seen in Fig.~\ref{fig1}b, for the structure ID of reflectance / transmittance spectra. Individual CNNs are trained independently for each leg of the inverse design triangle as shown in Fig.~\ref{fig1}a and independently for each total layer thickness, except the forward simulation (transfer matrix method) legs. This results in 20 individually trained networks (4 ID problems for each 1-5 layer system). Each CNN model architecture is individually optimized in terms of the total number of convolutional layers, dense layers, nodes per layer, dropout rate and a scaling factor of the learning decay rate. The variance in optimal CNN hyperparameters for different ID problems is plausible, since the design space can be drastically different between individual ID problems. The utilized network structure for each individual network can be found in the Supporting Information.

\subsection{Network Training and Evaluation}

CNN training is performed using Tensorflow and Keras software in a Python 3 environment. All models were trained for 300 epochs, with a set of 200,000 independent structures. The training learning rate decay is of the form $a/(\sqrt{t}+1)$, where $a$ is a scaling factor optimized for each individual network. All training was performed on a single high performance computing node with 10 cores and 64GB RAM, featuring an NVIDIA 2080 GPU. Full training of the CNN in this environment takes roughly three hours. The CNN evaluation was performed on an independent test set, consisting of 20,000 independent structures drawn from the same statistical range as the training set. For consistency, all networks and comparison optimization methods were evaluated on a high performance computing node with 24 cores and 64GB RAM on the same server and in the same environment as the network training. Comparison methods were fully parallelized to take full advantage of the available resources. Computational resource availability was consistent for all comparisons shown in this work. A full description of the python environment can be found in the Github repository associated with the paper (link given below).

\subsection{Loss Function}

The CNN training is guided by the loss function, which is a mathematical function providing a quantitative measure of the CNN efficiency by comparing the CNN output for known 'ground truth' examples. During training the CNN optimizes internal weight parameters by minimizing the total loss via the back-propagation algorithm. In structure ID, the loss function is the categorical cross-entropy of individual layer material predictions plus two times the mean squared error (MSE) in the predicted layer thicknesses. Note that layer thickness MSE is correlated with material loss, since the choice of predicted material in each layer necessarily influences the optimal thickness of that layer. In spectral ID, the training loss is simply the MSE in the output spectral response.

\begin{acknowledgement}
We acknowledge support from the Ohio Third Frontier Project "Research Cluster on Surfaces in Advanced Materials" (RC-SAM) at Case Western Reserve University. A.L., M.H. and G.S. acknowledge financial support from the NSF Grant no. 1904592 “Instrument Development: Multiplex Sensory Interfaces Between Photonic Nanostructures and Thin Film Ionic Liquids”. This work made use of the High Performance Computing Resource in the Core Facility for Advanced Research Computing at Case Western Reserve University.
\end{acknowledgement}

\begin{suppinfo}

Supporting information. Link to repository for source code, link to repository for CNN models, material parameters, dataset analysis, MSE landscapes, solution space analysis, CNN results analysis, impact of spectral features, details of the utilized CNN models, comparison with other machine learning methods, comparison with the needle method.

\end{suppinfo}

\bibliography{achemso-demo}




\end{document}